# Skills for forecasting space weather

**H. J. Austin[1] and N. P. Savani[2,3]**

[1]*Johns Hopkins University, Baltimore, MD, USA*
[2]*Heliospheric Division, NASA Goddard Space Flight Center, Greenbelt, MD, USA*
[3]*Baltimore County Goddard Planetary Heliophysics Institute, University of Maryland, Baltimore, MD, USA*

## Introduction

Space weather refers to the dynamic conditions surrounding the Earth and driven by the influence of the Sun's activity. Transient features emanating from an active Sun, such as solar flares, solar energetic particles and coronal mass ejections (CMEs), create drastic changes to the quasi-static space environment surrounding the Earth. CMEs are often identified as the source of severe geomagnetic storms that are observed at Earth (Tsurutani *et al.,* 1997; Green and Baker, 2015).

As national economies grow and become more reliant on technology, their vulnerability to space weather grows (Oughton *et al.,* 2017). As a result, the impact of space weather has become more prominent to an ever-growing audience (Gibbs, 2014). A number of nations have begun a coordinated approach to addressing these risks and forecasting the environment, for example in the USA (Lanzerotti, 2004; OSTP, 2015a), the UK (Gibbs, 2014), Japan (Ishii, 2017), and Mexico (Gonzalez-Esparza *et al.,* 2017). In 2012, the UK entered space weather behavior into their National Risk Register and placed it as the fourth highest risk.

The influence of space weather spreads further than just commercial entities and governmental policy, and sometimes reaches public awareness (Lanzerotti, 2011), for example through its effects on the tourism industry based around the Northern Lights (MacDonald *et al.,* 2015). There have also been several smartphone apps designed with the specific intention of enabling the general population to become more proficient in accessing and understanding space weather forecasting (Tobiska *et al.,* 2010).

A key objective of space weather forecasting is not only to improve the accuracy of our forecasts but also to improve the lead time in which accurate forecasts can be made. Many industrial sectors affected by space weather often desire a minimum of 24h lead time warning. Achieving these goals requires the scientific community to overcome several challenges.

CMEs, the major driver of near-Earth space weather effects, are often coherent structures of magnetised plasma (Burlaga, 1988). The Large Angle and Spectrometric COronagraph experiment (LASCO) on the SOlar and Heliospheric Observatory (SOHO) spacecraft is used to collect chronographic imagery of CMEs. Earth-directed events are often seen as partial halos or full halos, as shown in Figure 1, depending on the position of the source location. More centre-disc eruptions appear as full halos, while eruptions with larger-in-magnitude longitudes appear as partial halos.

After the initial eruption, forecasters who observe CMEs can map the event to Earth-arrival using several models, a procedure outlined in Figure 2. The orientation of the magnetic vectors within CMEs is a key influence on whether the energy and plasma transfer is enhanced or inhibited between the energetic CME and the Earth's magnetosphere, which often acts like a protective bubble around the globe (Dungey, 1961), see Figure 3. A southward-directed solar wind magnetic field is more likely to lead to magnetic reconnection, a process by which solar wind hits the Earth's magnetosphere, causing the field lines to break and recon-nect while releasing energy and causing geomagnetic storms.

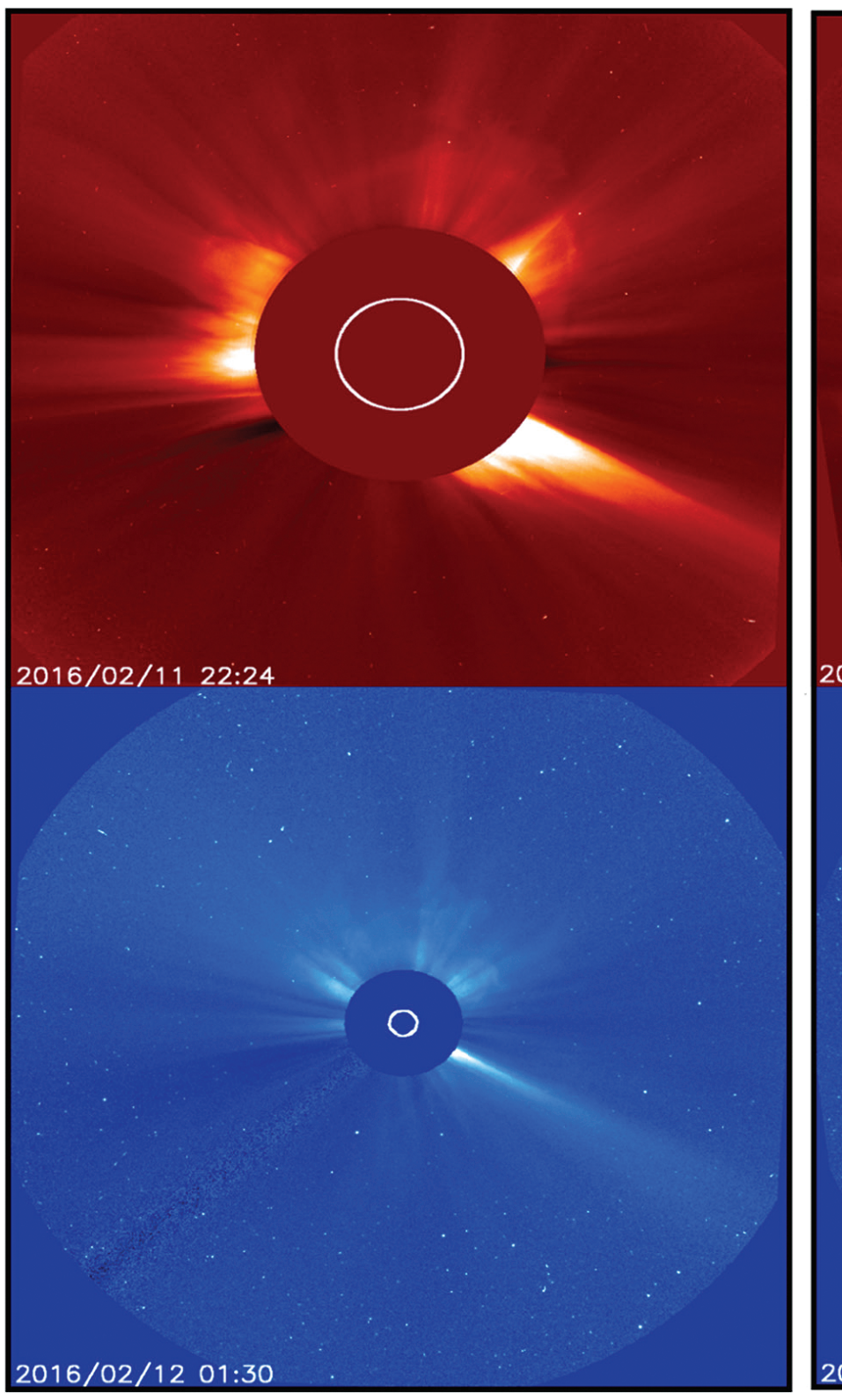

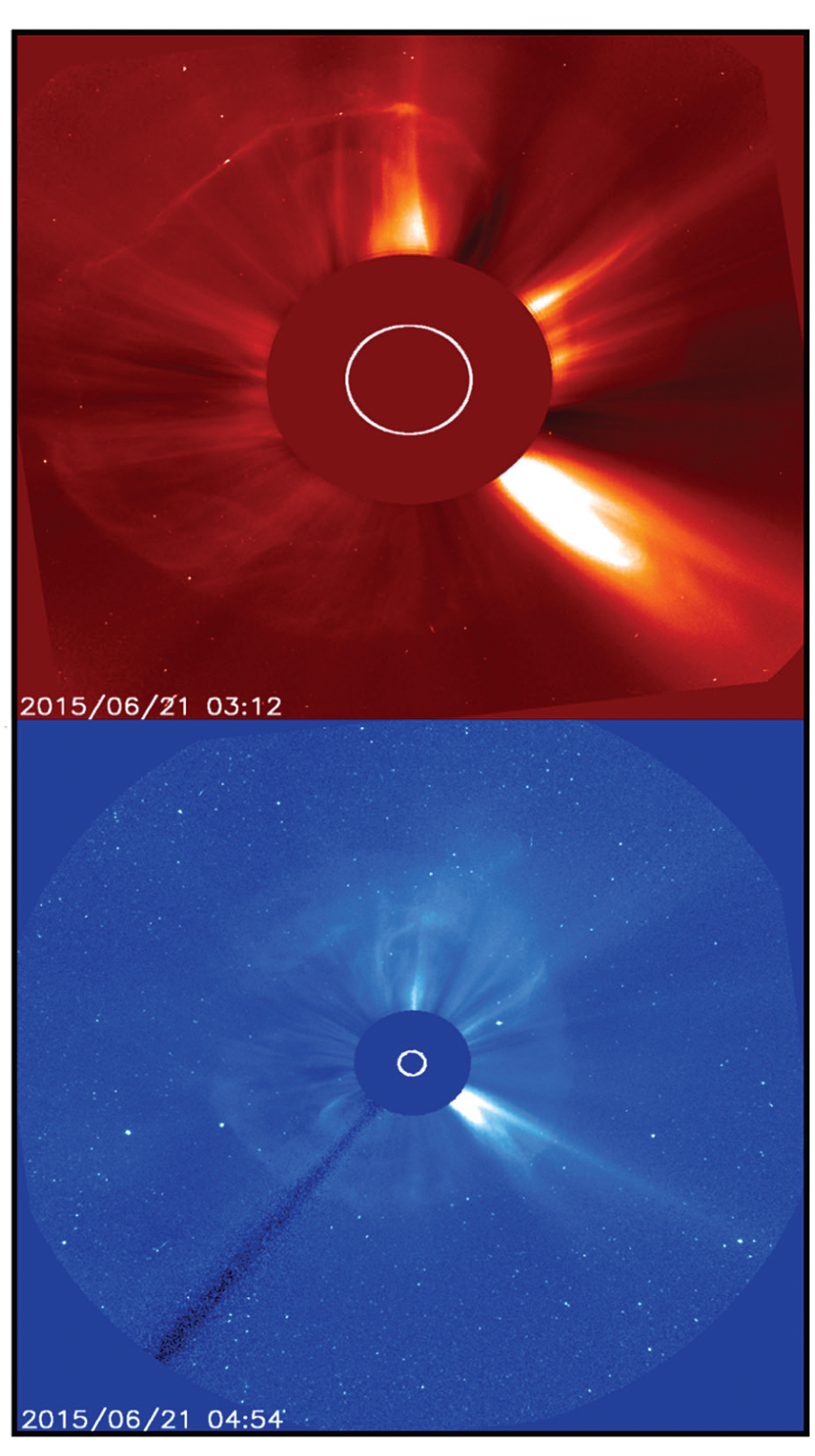

*Figure 1. Images of a partial halo CME (left) and full halo CME (right), shown using LASCO C2 (top images) and C3 (bottom images) instruments. A CME can be seen to propagate away from the Sun, first in C2 and then in the wider camera of C3. The bright light of the Sun (located in the central white line) is blocked by the instruments so that the darker and tenuous plasma of the CME can be observed. A halo CME is where the CME structure fully surrounds the occulting disc that blocks the Sun and is usually considered to be moving directly towards the Earth.*

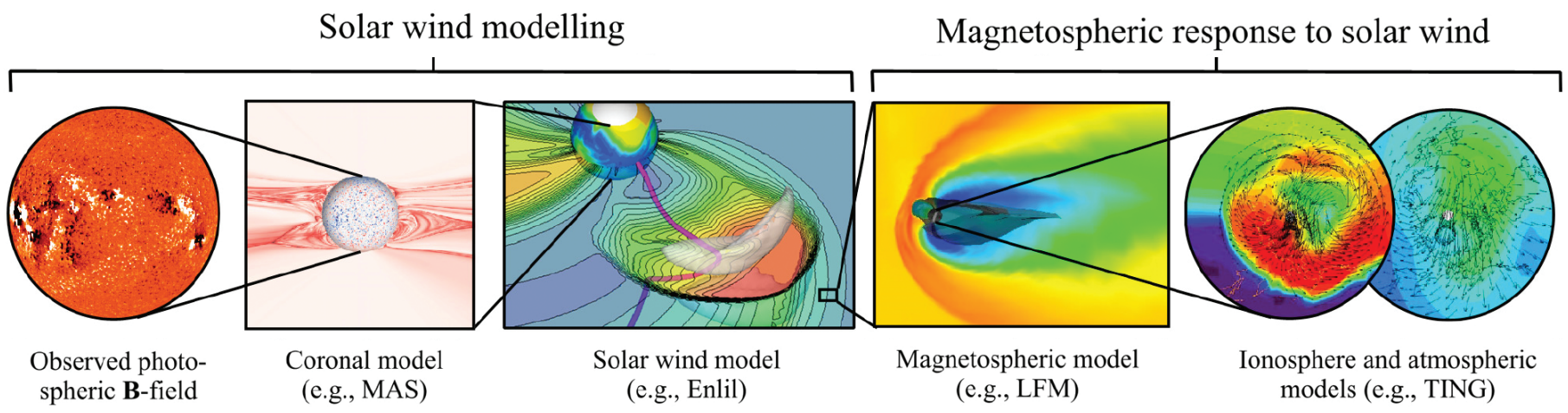

*Figure 2. A pathway of how numerical simulations can be coupled together from the solar surface to Earth. (Source: Owens et al., 2014).*

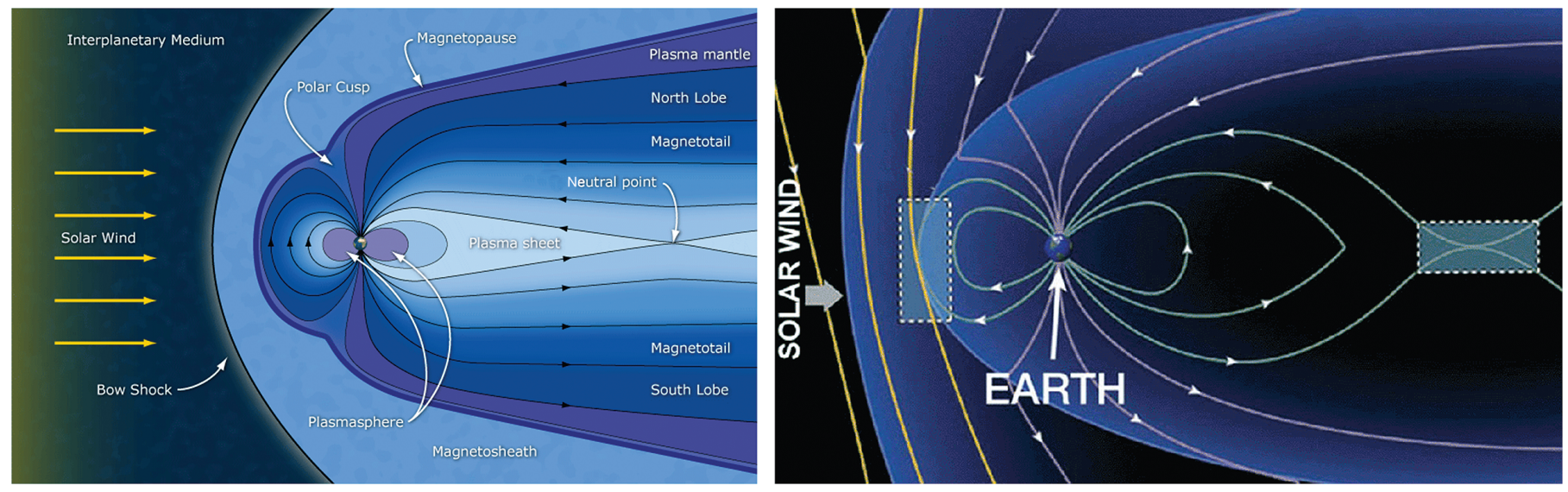

*Figure 3. Magnetosphere graphic (left) and reconnection schematic (right). Reconnection is a process that describes the combination of different magnetic field lines to create a new magnetic topology, shown as regions within the dashed box of the right panel. (Sources: European Space Agency (ESA) (left); National Aeronautics and Space Administration (NASA) (right).)*

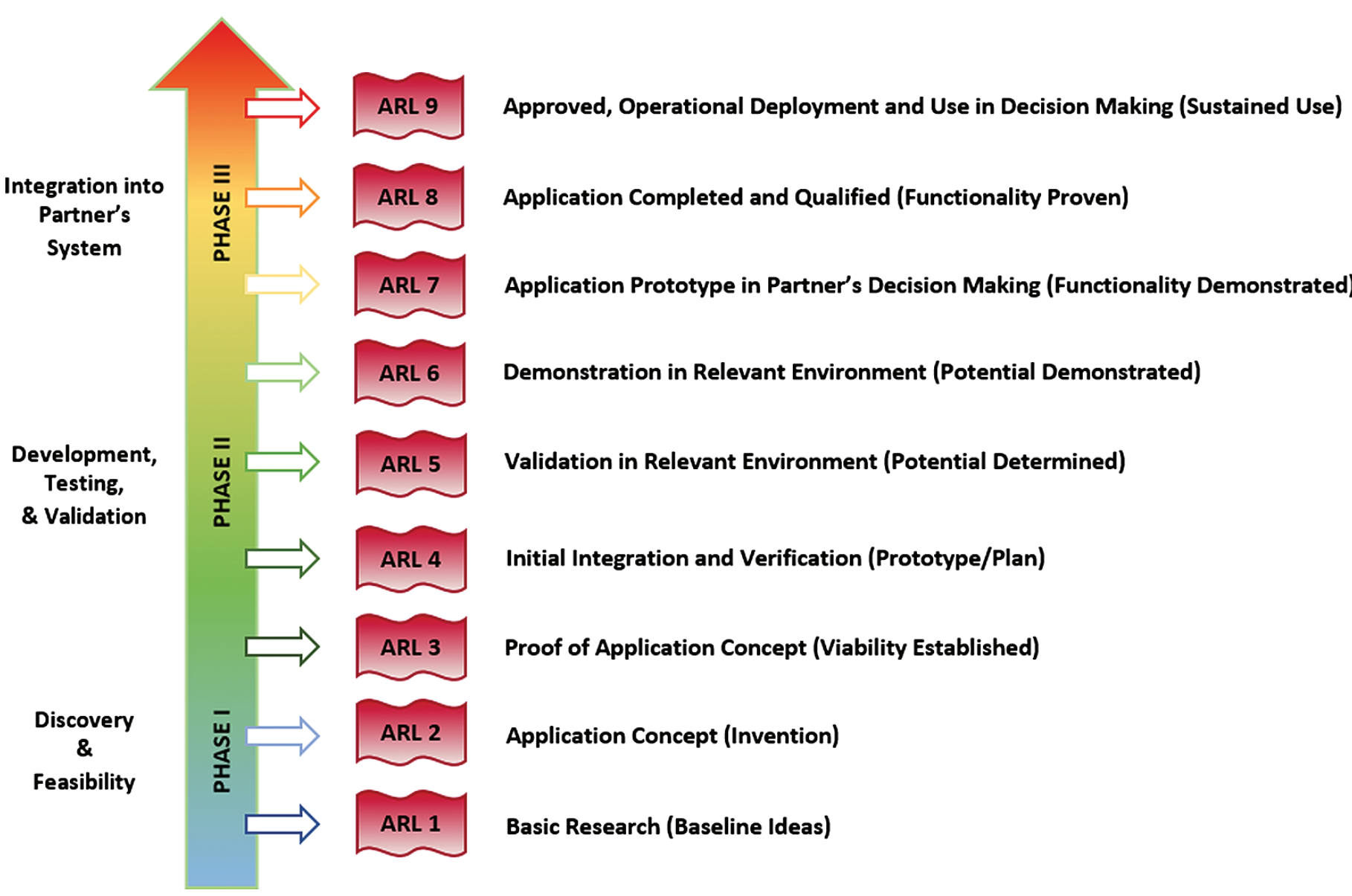

*Figure 4. Progress of research development into operational usage via Application Readiness Levels (ARLs) from the NASA applied science program. (Adapted from: Pulkkinen, 2017).*

Savani *et al.* (2015) first described a model architecture capable of predicting the magnetic vectors inside CMEs with a lead time of more than 24h. This architecture described a modular system that improved and incorporated several empirical relationships between the Sun and Earth environments. Eight test CMEs were investigated under this architecture, along with a design of evaluating the skill based on the National Oceanic and Atmospheric Administration (NOAA) Space Weather Prediction Center (SWPC) geomagnetic storm scale (G-scale; Savani *et al.*, 2017). The G-scale is taken from the geomagnetic (planetary) Kp-index, which is a quasi-logarithmic planetary average of geomagnetic activity measured over 3h synoptic periods.

In this article, we describe an overview of the process of moving a research idea into operational use and demonstrate not only a statistically reliable forecast system using analysis from near real-time events, but also the importance of generating accurate uncertainty limits.

## Research to Operations

US National Space Weather Strategy (OSTP, 2015a) has identified a structure for improving and preparing the nation for near- and long-term space weather effects. The National Space Weather Action Plan (OSTP, 2015b) supports these same goals with mandated actions and timelines, as well as coordinating the various United States departments and agencies.

As part of the strategy to improve forecasts, NOAA, on behalf of the National Science and Technology Council, has drafted a plan to improve and formalise the development of research capabilities into operational usage (Research to Operations (R2O); Jonas *et al.*, 2017). Contributing to this development, a pathway has been described for the coupling of R2O activity with Operations to Research (O2R). O2R describes the reverse process of R2O and, for example, includes work towards: (1) upgrading and enhancing existing operational models and products, and (2) testing and evaluating operational model performance by researchers for fundamental scientific discovery.

NASA's Community Coordinated Modelling Center (CCMC) space weather team uses the research-to-operation pathway implemented in Applied Sciences Programme for the Earth sciences. This pathway provides guidance of how specific research can become closer to fully operational and quantify the level of maturity it has achieved. More specifically, a nine-step Application Readiness Level (ARL) index has been implemented to track and manage the milestones for R2O. Figure 4 shows a summary of the Applications Readiness Level (ARL) pathway (Pulkkinen, 2017), which has been adapted from a scale used by NASA for managing technology development and risk (Technology Readiness Levels, TRL).

## Operation of Bz4Cast tool

The Bz4Cast tool is a simplified prototype space weather forecasting product (ARL 4) that encompasses the technique described by Savani *et al.* (2015, 2017). Figure 5 shows a summary of the modular architecture from the original technique, highlighting the ease with which more advanced capabilities can be added by switching and replacing various models and empirical relationships.

As part of the R2O activity, the Bz4Cast forecast system requires validation under realistic and relevant environmental conditions (ARL 5). In this study we validated the operational system through the use of previously measured CMEs that were studied internationally as part of ongoing operations scoreboard and skill testing.

## Event selection

The choice of CMEs investigated was based on information from the Space Weather Database Of Notifications, Knowledge, Information (DONKI), which was developed at the CCMC and hosted at NASA Goddard Space Flight Center (GSFC). The DONKI catalogue provides a comprehensive database of events investigated in a forecasting-style real- and near-real-time mode.

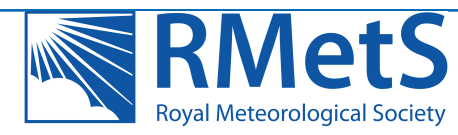

FR structure model + FR orientation @ 1AU model → Earth trajectory estimate

Chirality model + |B| field @ 1AU model → + → FR Predicted B vector field @ 1AU = Bz4Cast model

Sheath model → + → CME Predicted B vector field @ 1AU

*Figure 5. Synopsis of the modular architecture within the original Bz4Cast prediction model, along with an updated prototype that includes the shock and sheath region.*

The CME events in this paper were chosen based on the detection of interplanetary shocks from the DONKI catalogue between April 2010 and July 2016. A total of 53 events were found, based on the following criteria: (1) the source of the CME occurred between solar longitude −60 and 60°; (2) expected partial impacts for Earth were excluded when looking at the heliospheric ENLIL simulations; (3) the ENLIL timeline predicted a clear jump in Kp when it arrived, (4) the solar wind data at the Lagrangian 1 point (L1), a point located between the Sun–Earth line where spacecraft are in gravitational equilibrium, displayed a spike in the total magnetic field strength to indicate a clear arrival; and (5) there was only one CME contributing to the interplanetary shock (Bz4Cast system does not currently operate for multiple simultaneous CME events).

These events were then subjected to a finer search and were down-selected to 25 events based on the condition that their solar source region could be clearly identified using imagery from the solar Dynamics Observatory. This final selection of CME events is considered to be reliably Earth-directed with a clear interplanetary signature measured upstream of Earth at L1.

The CME measurements for longitude, latitude, half-width, and arrival time were pulled from DONKI. The latitude and longitude measurements were visually compared with SDO images to ensure accuracy. Arrival time was time-shifted to match the actual peak Kp values.

The final 25 events were investigated using the Bz4Cast forecast system in two different modes. The first method incorporates L1 solar wind predictions (maximum speed and maximum magnetic field strength value) from the ENLIL forecast simulation (called Bz4Cast-E). The second method used the actual measured values from L1 once the CME had arrived (called Bz4Cast-L1). Thus, method 2 tests whether the Bz4Cast technology is limited by the prediction accuracy of the near-Earth environment. Both Bz4Cast prediction methods were then compared with NOAA's 3-day forecasts.

It should be noted that ENLIL frequently underpredicts maximum magnetic field strength. This was observed for 23 out of 25 CMEs studied (and for 49 from 53 in the initial selection). This is a current limitation during real-time forecasting.

To compare the effectiveness of the Bz4Cast tool, skill scores were run on NOAA predictions of Kp and both Bz4Cast methods. Data was pulled from the earliest NOAA 3-day geomagnetic forecast following the eruption that predicted an arrival, occasionally spanning two forecasts if all eleven synoptic periods were not covered in the first report.

## Skill metrics

To evaluate the quality of the forecasting NOAA's G-scale, the Bz4Cast system converts predicted solar wind magnetic vectors into the G-scale (via the Kp index measured on Earth).

The skill component of the prediction is performed by isolating these field characteristics from the other forecasting elements that lead to additional uncertainty (that is, arrival time and velocity).

Therefore, to evaluate the predictions of any given coronal mass ejection (CME) on Earth, the predicted Kp values were all time-shifted to the observed Kp data. Independently for each CME, the size of the time difference was dictated by the location of a maximum correlation between the predicted and observed data.

This article's approach to skill takes advantage of methodologies that are more established in the terrestrial weather community. The focus of a skill evaluation is to assess the long lead time predictions of the solar wind field vectors and G-scale.

Skill scores are evaluated based on an event as defined by each Kp synoptic period. The skill criteria for the contingency table (hit, false positive, miss and correct null) are based on the G1 storm scale (Kp = 5). For a hit, a combined criterion was required, such that: (1) either prediction or observation recorded a Kp ≥ 5, and (2) the observed and predicted Kp were within 1.5 of each other.

For both trials of the Bz4Cast system and the 3-day NOAA forecast, each of the 25 events (each with 11 synoptic periods) were compared with actual Kp data. A contingency table for the combined 25 events is displayed in Table 1.

These contingency tables were used to estimate a variety of skill metrics, which are defined as follows:

$$\text{True skill statistic} = \frac{(AD-BC)}{\left[(A+C)\,(B+D)\right]} \quad (1)$$

$$\text{Proportion Correct} = \frac{(A+D)}{n} \quad (2)$$

$$\text{Hit Rate} = \frac{A}{(A+C)} \quad (3)$$

$$\text{Frequency Bias} = \frac{(A+B)}{(A+C)} \quad (4)$$

$$\text{Threat Score} = \frac{A}{(A+B+C)} \quad (5)$$

$$\text{False Alarm Ratio} = \frac{B}{(A+B)'} \quad (6)$$

where, $A$ = hit; $B$ = false alarm; $C$ = miss; $D$ = correct null. True skill statistic (TSS) compares correct predictions with the cases in which an event occurred, with values in the range −1 ≤ TSS ≤ 1. Proportion Correct (PC) shows the accuracy of the prediction, regardless of whether or not an event occurred, with values in the range 0 ≤ PC ≤ 1. Hit Rate (HR) can be interpreted as how often an event was correctly predicted, for the cases in which an event occurred, with values in the range 0 ≤ HR ≤ 1. Frequency Bias (FB) is the number of predicted events over the total number of observed events, with

**Table 1**

*Tally of each 3h Kp event for the 25 CMEs used for skill scores for NOAA data and both Bz4Cast methods. Bz4Cast-L1 integrates the actual interplanetary data from the L1 location; Bz4Cast-E integrates the estimated solar wind parameters from the real-time runs of ENLIL simulations performed by the CCMC at NASA GSFC.*

| NOAA | | | Bz4Cast-L1 | | | Bz4Cast-E | | |
|---|---|---|---|---|---|---|---|---|
| *Event forecast* | *Event observed* | | *Event forecast* | *Event observed* | | *Event forecast* | *Event observed* | |
| | *Yes* | *No* | | *Yes* | *No* | | *Yes* | *No* |
| Yes | 33 | 16 | Yes | 21 | 38 | Yes | 24 | 27 |
| No | 31 | 195 | No | 35 | 181 | No | 36 | 188 |

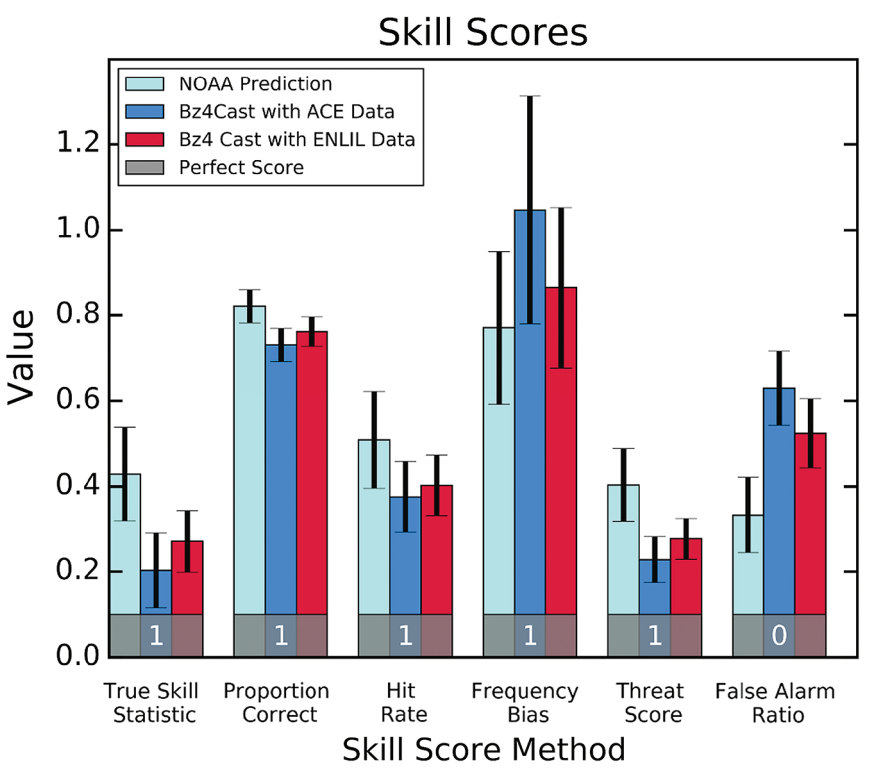

*Figure 6. Skill score results for all three methods, shown with the perfect score for each statistic.*

values in the range 0 ≤ FB ≤ ∞. Threat Score (TS) is the number of correctly predicted events over the total number of events (predicted and observed), with values in the range 0 ≤ TS ≤ 1. This Threat Score (TS) looks only at situations that might or did produce an event and compares it with how well the tool predicted an outcome. False Alarm Ratio (FAR) is the number of false events out of the number of predicted events, with values in the range 0 ≤ FAR ≤ 1.

Figure 6 shows a comparison between the skill metrics of the NOAA 3-day forecast and the Bz4Cast technology.

## Discussion

A comparison of skill between the Bz4Cast-E and Bz4Cast-L1 indicates there is a slightly lower Proportion Correct (PC) and Hit Rate for the method using the real solar wind data. The performance of Bz4Cast-L1 demonstrated a slightly higher False Alarm Ratio (FAR), which was expected as ENLIL often underpredicts $B_{max}$ (the maximum magnetic field during the passage of a CME). By increasing $B_{max}$ from the ENLIL values to the L1 data, the prediction of maximum Kp and G-scale also increases. The Bz4Cast-L1 results also had the closest ideal Frequency Bias (FB) estimate; however, the performance was lower under the evaluation of Threat Score and TSS.

In general, the two Bz4Cast methods produced similar results within error limitations. However, a larger sample size is desirable for the performance of more reliable hypothesis testing to estimate the extent to which the different methods generate different populations of skill.

Without looking at the uncertainty in the results, NOAA outperformed the Bz4Cast runs on five of the skill metrics, while Bz4Cast only outperformed NOAA in terms of Frequency Bias. This is because Bz4Cast tends to produce more false positives than NOAA's prediction; it also produces less hits and correct nulls but similar numbers of misses. However, when we consider the uncertainty in the skill of each forecast method, they demonstrate equal performance under a single standard deviation of the uncertainty. The uncertainty in each metric was calculated with a bootstrapping methodology whereby resampling was performed using a Monte Carlo algorithm under a sampling with replacement scenario.

**Table 2**

*A comparison of the number of events counted for each method using the Kp ≥ 5 (as used in the skill scores) and Kp ≥ 6 (which is the threshold used by the CCMC for alert purposes) definitions of an event.*

| | *Times Kp ≥ 5* | *Percent of times Kp ≥ 5* | *Times Kp ≥ 6* | *Percent of times Kp ≥ 6* |
|---|---|---|---|---|
| NOAA | 40 | 14.6% | 14 | 5.1% |
| Bz4Cast with ACE | 86 | 31.3% | 68 | 24.7% |
| Bz4Cast with ENLIL | 86 | 31.3% | 53 | 19.3% |
| Actual | 54 | 19.6% | 27 | 9.8% |

During the testing and analysis of the Bz4Cast methodology, many of the input parameters of the CME origins were tested through user experience. This was to complement the more theoretical study of the parameters performed by Savani *et al.* (2017). CME source parameters such as tilt, size and location were measured. However, to better emulate the practical approach of using the experience of an on-duty forecaster, freedom to make small adjustments to the parameters was given. To ensure skill remained unbiased by the final metrics, any adjustments were made and fixed prior to investigating the skill.

Of the 275 synoptic periods studied, NOAA predicted that 5.1% of them resulted in a Kp ≥ 6, while 9.8% of the time Kp actually reached or surpassed 6. For the same synoptic periods, both trials of Bz4Cast overpredicted Kp, using ACE data (24.7%) and ENLIL data (19.3%) (Table 2).

Bz4Cast performed equally well or better than NOAA on events with higher Kp values (≥7), though these events are much rarer (these values occurred during four CME events, on 28 September 2012, 10 September 2014, 15 March 2015 and 21 June 2015). NOAA performed better than Bz4Cast in cases where Kp = 6, as Bz4Cast tended to predict larger storms than actually occurred. Of the 25 events that were studied, 13 produced geomagnetic storms of Kp ≤ 5, and as NOAA tended to predict smaller storms, these errors arise as a result of cases in which there is an actual Kp of 5 and a NOAA prediction of 4; this would still count as a hit, but NOAA will not have not successfully predicted an event. This is also highlighted in the underprediction of NOAA results shown in the Frequency Bias.

## Conclusions

In this paper, we have broadened the statistical verification of the first empirically-driven model to forecast the solar wind magnetic vectors inside a CME prior to their arrival at Earth. A total of 25 CME events were tested with the Bz4Cast model, and the forecast skill was tested against NOAA's SWPC G-scale for geomagnetic storms.

Future research will also focus on bolstering the statistics by increasing the number of events. However, to produce statistics for approximately 100 events, historical datasets will be required such that direct comparisons of the same events to the actual NOAA forecasts using their methodology will not be possible. Thus, the direct comparative study in this paper serves as a testbed that can provide a benchmark for how later studies of the model can be statistically validated under more real-time and operational scenarios.

Early indications of this Bz4Cast forecast model suggest that a slightly higher false alarm rate may become persistent, while NOAA 3-day forecasts might be slightly more biased towards underpredicting. For end users such as SKYNET spacecraft management authority at Airbus, a slightly elevated false alarm rate for an early warning system is potentially preferable to the potential losses of a miss scenario (Haggarty, 2017). Due to a high emphasis on reliability of service, potential long lead time warnings enable end users to start risk mitigating procedures, which can be less detrimental than an unexpected loss to their service.

## Acknowledgements

We would like to thank the CCMC hosted at NASA Goddard Space Flight Center and The Catholic University of America's Scientific and Engineering Student Internship programme for their support and guidance. NPS thanks the resources provided by TEDCO Maryland Innovation Initiative. We would also like to thank Collin Van Son for his initial work with the Bz4Cast tool.

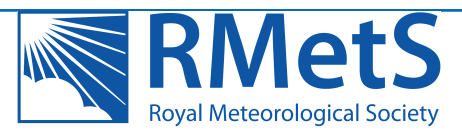

*Correspondence to: H. J. Austin*

*haustin1@jhu.edu*

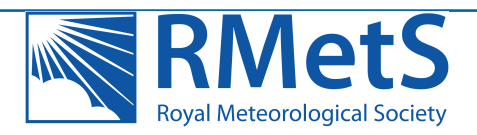